\documentstyle[aps,preprint]{revtex}  
\newcommand{\vsig}{\mbox{\boldmath $\sigma$ \unboldmath}} 
\newcommand{\veps}{\mbox{\boldmath $\epsilon$ \unboldmath}}  
  
\newcommand{\vtau}{\mbox{\boldmath $\tau$ \unboldmath}}  
\newcommand{\vpi}{\mbox{\boldmath $\pi$ \unboldmath}}  

\begin{document}  
\bibliographystyle{unsrt}  
  
\title{\bf Vector Meson Photoproduction with an Effective
Lagrangian in the Quark Model II: $\omega$ Photoproduction}  
  
\author{Qiang Zhao, Zhenping Li \\  
Department of Physics, Peking University, Beijing, 100871, P.R.China\\
C. Bennhold\\
Department of Physics, Center for Nuclear Studies,\\
 The George Washington University, Washington, D.C., 20052, USA}  
\maketitle  
  
\begin{abstract}  
An investigation of $\gamma p\to \omega p$  is presented in a constituent
quark model approach.  The sparse data in the large $t$ region where
the resonances dominate is well described within the model,
while the diffractive 
behavior in the small $t$ region requires an additional t-channel exchange. 
Taking into
 account the t-channel $\pi^0$ exchange, we find a good 
overall agreement with the available data with only 3 free parameters. 
Our study shows that the differential cross section is not sensitive to
s-channel resonances, however, the polarization observables are demonstrated
to be very sensitive. Thus, measuring polarization observables
is the crucial part of the vector meson 
photoproduction program in the search for ``missing" resonances.
\end{abstract} 

\section*{\bf 1. Introduction}  
The newly established electron and photon facilities have made it possible to 
investigate the mechanism of vector meson 
photoproduction
 on nucleons with much improved experimental accuracy. This has been
motivated in part by 
the puzzle that the NRQM ~\cite{NRQM,capstick} predicts a much richer resonance
 spectrum than has been observed in $\pi N\to \pi N$ scattering
experiments. 
Quark model studies have suggested that those resonances missing in the
$\pi N$ 
channel may couple strongly to, i.e., the $\omega N$ and $\rho N$
channels. 
Experiments have been performed at ELSA~\cite{elsa} and will be done 
at TJNAF in the near future~\cite{cebaf}. 
Therefore, a theory on the reaction mechanism 
that highlights the dynamical role of 
s-channel resonances is crucial in order to interpret new 
experimental data.

Building on recent successes of the quark model 
approach to pseudoscalar meson 
photoproduction we have proposed\cite{vector} extending this 
approach to vector meson 
photoproduction.  The focus of this paper is to present the numerical
implementation of the
 reaction $\gamma p\to \omega p$ in the quark model framework. 
The reaction $\gamma p\to \omega p$ shares similar features with the
reaction
$\gamma p \to \eta p$ in the pseudoscalar sector, since 
both are isospin zero states. 
This eliminates contributions from all isospin 3/2 intermediate
resonances, thus the signals from
isospin 1/2 states, in particular those resonances ``missing" in
the $\pi N$ 
channels, can be isolated . 
Thus, investigating the dynamical role of the 
s-channel resonances  and examining which
experimental observables are best suited to demonstrate 
the existence of these resonances is the
primary goal of 
our study.  However, an important question related to this discussion 
 is whether additional t-channel exchanges responsible
for the diffractive behavior 
in the small $t$ region are needed. 
It has been proposed in the pseudoscalar sector that the extra t-channel 
exchanges could be excluded with the duality 
hypothesis\cite{duality}. In a study of
 kaon photoproduction by 
J.C. David {\sl et al}~\cite{david}, the inclusion of the additional
 s-, u-channel resonances, particularly the higher partial wave resonances 
leads to smaller coupling constants for the t-channel $K^*$ exchanges,
indicating some phenomenological support for 
the duality argument. Similarly, one could also
 investigate the role of the t-channel
 exchanges in $\omega$ photoproduction by including the t-channel
exchanges and
 treating the coupling constants as parameters. 
If the diffractive behavior could be 
described by contributions from a complete set of s- 
and u-channel resonances, the coupling constants for the 
t-channel exchanges should
 be very small.  Thus, we include in our calculations the $\pi^0$ exchange 
suggested by Friman and Soyeur~\cite{FrimanSoyeur}, 
who showed that the diffractive behavior
  in the low energy region
can be well described through $\pi^0$ exchange.

In section 2, we give a brief description of our model, of which
 a complete framework has been given in Ref.~\cite{vector}. The $\pi^0$
exchange model 
by Friman and Soyeur will be introduced in the standard helicity representation.  The
results of our calculations for the s- and u-channel transition  amplitudes are given in section 
3, and the conclusion are given in Section 4.

\section*{\bf 2. The model}
The starting point of our quark model approach to vector meson photoproduction is the 
effective Lagrangian~\cite{vector},
\begin{equation} 
\label{3.0}  
L_{eff}=-\overline{\psi}\gamma_\mu p^\mu\psi+\overline{\psi}\gamma_
\mu e_qA^\mu\psi +\overline{\psi}(a\gamma_\mu +  
\frac{ib\sigma_{\mu\nu}q^\nu}{2m_q}) \phi^\mu_m \psi,
\end{equation}
where $\psi$ and $\phi^\mu_m$ are the quark and the vector meson fields, and $a$ and $b$ are 
coupling constants which will be determined by experimental data.
At the tree level, the transition matrix element based on this   
Lagrangian in Eq.(\ref{3.0}) can be written as the sum of contributions  
 from s-, u- and t- channels,  
\begin{equation}  
M_{fi}=M^s_{fi}+M^u_{fi}+M^t_{fi}. 
\label{3.1}  
\end{equation}  
The first and second terms in Eq.(\ref{3.1}) represent the s- and u-channel
contributions, and a complete set of helicity amplitudes for each
of the s-channel resonance below 2 GeV have been evaluated
 in the $SU(6)\otimes O(3)$
 symmetry limit~\cite{vector}. Resonances above 2 GeV
are treated as degenerate in order to express contributions from  
resonances with a certain quantum number $n$  in a   
compact form.   The contributions from   
the u-channel resonances are divided into two parts as well. The   
first part contains the contributions from the resonances with   
the quantum number $n=0$, which includes the spin 1/2 baryons,  
 such as the $\Lambda$, $\Sigma$ and the nucleons, and the spin   
3/2 resonances, such as the $\Sigma^*$ in $K^*$ photoproduction   
and the $\Delta(1232)$ resonance in $\rho$ photoproduction. Because   
the mass splitting between the spin 1/2 and spin 3/2 resonances   
with $n=0$ is significant, they have to be treated separately.   
Since in $\omega$ photoproduction isospin 3/2 resonance 
cannot contribute due to
 isospin conservation, construction of the resonance contribution is simpler than 
in the case of $\rho$ meson photoproduction. The second part comes from the excited   
resonances with quantum number $n\ge 1$. Since the contributions  
 from the u-channel resonances are not sensitive to the precise 
mass positions, they are treated as degenerate as well.  The transition matrix elements are written 
in the form of helicity amplitudes, which are 12 independent amplitudes and defined 
as in ref.\cite{vector,tabakin},
\begin{eqnarray} \label{helicity}
H_{1\lambda_V}&=  &\langle \lambda_V, \lambda_2=+1/2|T|\lambda=1,   
\lambda_1=-1/2\rangle\nonumber\\  
H_{2\lambda_V}&=  &\langle \lambda_V, \lambda_2=+1/2|T|\lambda=1,   
\lambda_1=+1/2\rangle\nonumber\\  
H_{3\lambda_V}&=  &\langle \lambda_V, \lambda_2=-1/2|T|\lambda=1,   
\lambda_1=-1/2\rangle\nonumber\\  
H_{4\lambda_V}&=  &\langle \lambda_V, \lambda_2=-1/2|T|\lambda=1,  
\lambda_1=+1/2\rangle,  
\end{eqnarray}  
where the helicity states are denoted by $\lambda=\pm 1$ for the incident photon,   
$\lambda_V=0,\pm 1$ for the outgoing  vector meson, and $\lambda_1=\pm 1/2$, $\lambda_2=\pm 1/2$  
for the initial and final state nucleons, respectively.  The various experimental observables, 
such as the cross section and polarization observables,
have been discussed in
Ref.\cite{tabakin} in terms of these amplitudes.
   In particular, the differential cross section can be written as
\begin{equation}
\frac{d\sigma}{d\Omega_{c.m.}}=\frac{\alpha_e \omega_m(E_f+M_N)(E_i+M_N)}{8\pi s}
{|{\bf q}|}\frac 12
\sum^4_{a=1}\sum_{\lambda_V=0,\pm 1} |H_{a\lambda_V}|^2
\end{equation}
in the center of mass frame, where $\sqrt{s}$ is the total energy of the
 system, $M_N$ represents the mass of the nucleon, 
 $\omega_m$ denotes the energy of the meson with momentum ${\bf q}$,
 and $E_i$, $E_f$
denote the energies of the initial and final nucleon states. 
As defined in Ref.~\cite{tabakin}, the four single-spin observables in 
{\sl bilinear helicity product} (BHP) form have the following expressions:
\begin{equation}
\check{T}=-\frac{1}{2}\langle H|\fbox{$\Gamma^{10}$}\fbox{$\omega^1$}|H\rangle,
\end{equation}
for the target polarization,
\begin{equation}
\check{P}_{N^\prime}=\frac{1}{2}\langle H|\fbox{$\Gamma^{12}$}\fbox{$\omega^1$}|H\rangle,
\end{equation}
for the polarization of the final nucleon, 
\begin{equation}
\check{\Sigma}=\frac{1}{2}\langle H|\Gamma^4\omega^A|H\rangle,
\end{equation}
for the polarized photon asymmetry, and
\begin{equation}
\check{P}_V=\frac{1}{2}\langle H|\fbox{$\Gamma^1$}\fbox{$\omega^3$}|H\rangle,
\end{equation}
for the polarization of the final vector meson, 
where the explicit expressions and conventions for 
the $\Gamma$ and $\omega$ matrices have been given in Ref.~\cite{tabakin}.
The phase space factor for these four observables are the same as in
the differential cross section, therefore, 
they are normalized by being divided by the differential cross section. 

The t-channel exchange of $M^t_{fi}$ in Eq.(\ref{3.1}) would correspond to  $\omega$
exchange which is absent since the photon cannot couple to the $\omega$.
However, an additional t-channel exchange is commonly included even though
it is not a part of the Lagrangian in Eq.(\ref{3.0}).
This is the $\pi^0$ exchange
which is known to be needed in order to describe 
 the diffractive behavior in the small $t$ region.
The Lagrangian for the 
$\pi^0$ exchange model has the following form~\cite{FrimanSoyeur},
\begin{equation}\label{3}
L_{\pi NN}=-i g_{\pi NN}\overline\psi \gamma_5(\vtau\cdot\vpi)\psi
\end{equation}
for the $\pi NN$ coupling vertex, and 
\begin{equation}\label{4}
L_{\omega \pi^0 \gamma}=e_N\frac{ g_{\omega\pi\gamma} }{M_\omega}
\epsilon_{\alpha\beta\gamma\delta}\partial^\alpha A^\beta
\partial^\gamma\omega^\delta\pi^0
\end{equation}
for the $\omega\pi\gamma$ coupling vertex, where the $\omega^\delta$ and $\pi^0$  represent
 the $\omega$ and $\pi^0$ fields,  the $A^\beta$ denotes the electromagnetic field, 
and  $\epsilon_{\alpha\beta\gamma\delta}$ is the Levi-Civita tensor, and $M_\omega$ 
is the mass of $\omega$ meson. The $ g_{\pi NN}$ and $ g_{\omega\pi\gamma}$ in Eqs. 
(\ref{3}) and (\ref{4}) denote the coupling constants at the two
 vertices, respectively. Therefore, the transition amplitudes of t-channel $\pi^0$ 
exchange have the following expression,
\begin{equation}
M^t_T= \frac{e_Ng_{\pi NN} g_{\omega\pi\gamma}}{2M_\omega(t-m^2_\pi)}
\{\omega\veps\cdot({\bf q}\times\veps_v)+\omega_m{\bf k}\cdot(\veps\times\veps_m)\}
\vsig\cdot {\bf A}
e^{-\frac {({\bf q}-{\bf k})^2}{6\alpha_\pi^2}}
\label{t}
\end{equation}
for the transverse transition, and
\begin{equation}
M^t_L= -\frac{e_Ng_{\pi NN} g_{\omega\pi\gamma}}{2M_\omega(t-m^2_\pi)}
\frac{ M_\omega}{|{\bf q}|}(\veps\times{\bf k})\cdot{\bf q} \vsig\cdot {\bf A}
e^{-\frac {({\bf q}-{\bf k})^2}{6\alpha_\pi^2}}
\label{l}
\end{equation}
for the longitudinal transition, where
$\omega$ in the transition amplitudes
denotes the energy of the photon with momentum ${\bf k}$, and
${\bf A}=-\frac{{\bf q}}{E_f+M_N}+\frac{{\bf k}}{E_i+M_N}$,
and $t=(q-k)^2=M_\omega^2-2k\cdot q$.
The factor $e^{-\frac {({\bf q}-{\bf k})^2}{6\alpha_\pi^2}}$ in Eqs. 
(\ref{t}) and (\ref{l}) is the form factor for both $\pi NN$ and $\omega \gamma \pi$ vertices, if we assume
that the wavefunctions for nucleon, $\omega$ and $\pi$ have a Gaussian form.  The constant 
$\alpha_\pi^2$ in this form factor is treated as a parameter.   Following the same procedure
as in Ref.\cite{vector}, the explicit expressions for the operators in terms of the helicity 
amplitudes can be obtained. They are listed in Tables 1 and 2 for the transverse and 
longitudinal amplitudes respectively.

\section*{\bf 3. Results and Discussion}

Before discussing the details of our numerical results, we point out that the nonrelativistic 
wavefunction in the quark model become more inadequate as the energy of the system
 increases.  A procedure to partly remedy this problem is to introduce the Lorentz 
boost factor in the spatial integrals that involve the spatial wavefunctions of nucleons 
and baryon resonances,
\begin{equation}
R(q,k)\to \gamma_q\gamma_k R(q\gamma_q, k\gamma_k),
\end{equation}
where $\gamma_q=\frac{M_f}{E_f}$ and $\gamma_k=\frac{M_i}{E_i}$.  A similar procedure had
been used in the numerical evaluation of pseudoscalar meson photoproduction\cite{eta}.

There are four free parameters in the quark model approach to the s- and u-channel 
resonance contributions:  the quark mass $m_q$, the harmonic oscillator strength $\alpha$, 
and the coupling constants $a$ and $b^\prime=b-a$ from Eq. (\ref{3.0}). 
Because the quark mass $m_q$ and the parameter $\alpha$ are commonly used in 
the quark model, they are fixed at
\begin{eqnarray}
m_q&=&330 \ \mbox{MeV}\nonumber\\
\alpha&=&410\ \mbox{MeV}.
\end{eqnarray} 
In addition to the free parameters in the s- and u- channels, there are also parameters 
in the t-channel $\pi^0$ exchange: the coupling constants for the $\pi NN$ and $\omega \pi\gamma$
vertices, $g_{\pi NN}$ and $g_{\omega\pi\gamma}$, and the parameter $\alpha_{\pi}$  in Eqs. (\ref{t})
and (\ref{l}).  We find that the s- and u-channel resonance contributions 
alone are unable  to describe the  diffractive behavior in the small $t$ region.
We therefore include the $\pi^0$ exchange using for the coupling constants
$g_{\pi NN}$ and $g_{\omega\pi\gamma}$\cite{FrimanSoyeur}:
\begin{eqnarray}
\frac{g^2_{\pi NN}}{4\pi}&=& 14,\nonumber\\
g^2_{\omega\pi\gamma}&=&3.315,
\end{eqnarray}
and find a good  description of the diffractive behavior.  Note that the 
values of $g_{\pi NN}$ and $g_{\omega\pi\gamma}$ were fixed by separate experiments 
and, therefore, are not free parameters in Ref. \cite{FrimanSoyeur}.  
This suggest that the duality hypothesis may not work in vector meson photoproduction.
The parameter $\alpha_{\pi}$ will be determined from $\omega$
photoproduction data 
along with the coupling constants $a$ and $b^\prime$. Qualitatively, we would
expect that $\alpha_{\pi}$ be smaller than the parameter $\alpha=410$ MeV, since it represents 
the combined form factors for both $\pi NN$ and $\omega\pi\gamma$ vertices while the parameter
$\alpha$ only corresponds to the form factor for the $\pi NN$ or $\omega NN$ vertex alone.

In Table 3, we list the s-channel resonances that contribute to the $\omega$ photoproduction
in $SU(6)\otimes O(3)$ symmetry limit. The masses and widths of these resonances come from the 
recent Particle Data Group\cite{pdg}. It should be noted that the Moorhouse selection\cite{moor} 
rule have eliminated the states belonging to $[70, 1^-]_1$ and $[70, 2^+]_2$ representation with 
symmetric spin structure from contributing to the $\omega$ photoproduction with the proton
target so that the s-channel states $S_{11}(1650), D_{13}(1700), D_{15}(1650)$ are not present 
in our numerical evaluations. Of course, configuration mixing will lead to additional
contributions from these resonances  which, however, cannot  be determined at present due 
to the poor quality of data.

In Table 3 we also list the partial widths of the resonances 
decaying into the $\omega N$ channel and their photon decay 
hecility amplitudes. 
In terms of the helicity amplitudes, the partial width $\Gamma_\omega$ of a resonance with spin $J$ 
decaying into an $\omega$ meson and a nucleon is 
\begin{equation}
\Gamma_\omega=\frac{|{\bf q}|\omega_m}{4\pi}\frac{M_N}{M_R}\frac{8}{2J+1}
\{|A_{\frac 12}|^2+|A_{\frac 32}|^2+\frac{M^2_\omega}{|{\bf q}|^2}|S_{\frac 12}|^2 \},
\end{equation}
where $ A_{\frac 12}$, $ A_{\frac 32}$ and $ S_{\frac 12}$ represent the vector
meson helicity amplitudes and $M_R$ denotes the mass of the intermediate 
resonance.
The partial 
width $\Gamma_\gamma$ of the resonances decaying into $\gamma N$, 
and contributing to $\omega$ photoproduction, is 
\begin{equation}
\label{photonwidth}
\Gamma_\gamma=\frac{{\bf k}^2}{\pi}\frac{2M_N}{(2J+1)M_R}
\{|A^\gamma_{\frac 12}|^2+|A^\gamma_{\frac 32}|^2\},
\end{equation}
where $ A^\gamma_{\frac 12}$ and $ A^\gamma_{\frac 32}$ denote the photon helicity amplitudes.

Only the resonances $P_{13}(1900)$ and
$F_{15}(2000)$, at present classified as 2-star resonances
in the 1996 PDG listings, have masses above the $\omega$ decay threshold, 
and therefore have branching ratio into the $\omega N$ channel. We find that 
the $F_{15}(2000)$ has a larger decay width than the $P_{13}(1900)$.
This finding differs from Ref.~\cite{capstick}, 
where the $P_{13}(1900)$ was calculated to have the larger
width into the $\omega N$ channel. The photon decay helicity amplitudes 
in Table 3 are consistent with previous theoretical results 
of the NRQM approach in Ref.~\cite{NRQM}. Qualitatively, the resonance 
$F_{15}(2000)$ plays a very important role in 
$\omega$ photoproduction. Of course, since our investigation 
here is exploratory it
can only provide an approximate description of the 
resonance contributions; a more accurate approach to 
the intermediate resonance decay should be adopted in a
systematic analysis~\cite{vector}.

We have not performed a rigorous numerical fit to the available data 
because of the poor quality of the data.  
Our study suggests that the parameters $a$, $b^\prime$ 
and $\alpha_{\pi}$ with the values
\begin{eqnarray}
a&=&-2.2, \nonumber\\
b^\prime&=&3.0, \nonumber \\
\alpha_\pi & =  & 300 \ \mbox{MeV}
\end{eqnarray}
gives a good description of the differential cross section 
data~\cite{elsa} in the resonance region. Clearly, these
parameters have considerable uncertainties.

Fig. 1 shows our calculations for the differential cross section
at the average photon energies of  $E_\gamma=$1.225, 1.45, 1.675 
and 1.915 GeV, 
in comparison with the data~\cite{elsa}. The results for
the t-channel $\pi^0$ exchange and contributions from only the s- and u-channel processes 
are also shown separately.
Our results with the $\pi^0$ exchange are consistent with the findings of
Ref. \cite{FrimanSoyeur} although the form factor in our calculation is different.
Fig. 1 clearly demonstrates that
the t-channel $\pi^0$ exchange is dominant in the small angle region, while the s- and u-channel
resonance contributions become more important as the scattering
 angle $\theta$ increases. 
To test the sensitivity of s-channel resonances to the differential 
cross section, the angular distribution at 1.675 GeV is presented 
with and without the contribution from the resonance $F_{15}(2000)$;
its threshold energy is around 1.675 GeV in the lab. frame. The results indicate that 
the differential cross section data alone are not sufficient to 
determine the presence of this resonance considering the theoretical 
and experimental uncertainties. Since the resonance couplings of the
$F_{15}(2000)$ are larger than those of other resonances in this 
mass region, the sensitivity of the differential cross section to
other resonances around 2 GeV is even smaller. 

In contrast to the differential cross section, 
the polarization observables show a much more dramatic dependence 
on the presence of the s-channel resonances. We present
results of four single polarizations at 1.675 GeV in Fig. 2. The absence 
of the resonance $F_{15}(2000)$ leads to a sign change in the 
target polarization, and the variations in the recoil as well 
as the meson polarization observable are very significant as well.
The absence of the resonance 
$P_{13}(1900)$, also shown in Fig. 2,
leads to very significant changes in the recoil 
polarization. Although we 
do not expect our numerical results to give a quantitative
prediction of
polarization observables at the present stage, since the calculations are limited 
to the $SU(6)\otimes O(3)$ symmetry limit that should be broken in 
more realistic quark model wavefunction, our results clearly suggest that the 
polarization observables may be the best place to determine s-channel 
resonance properties.

Our results for the total cross section are shown in Fig. 3, 
in which the contributions from the s- and
u-channel resonances alone are compared to the full calculation. 
Our results indicate
an increasing discrepancy between theory and 
the data~\cite{elsa,ABHMC,data}
with increasing energy $E_{\gamma}$,
This discrepancy 
comes mainly from the small angle region where the $\pi^0$ 
exchange alone is not sufficient to describe the
diffractive behavior at higher energies.  One might expect 
that  Pomeron exchange\cite{pomeron,wolf}
plays a more important role in the higher energy region.  
However,  Fig. 1 shows that our results for 
the differential cross section at the large angle region are 
in good agreement with the data, and it 
suggests that contributions from the s- and u- channel resonances
which are the main focus of our study, give
an appropriate description of the reaction mechanism. 

It is interesting to note that the small bump around 1.7 GeV in the total cross section
comes from the contributions of the resonance $F_{15}(2000)$.
As discussed above, our calculations
find that the resonance $F_{15}(2000)$
has a strong coupling to the $\omega$ N channel. Thus, 
this resonance is perhaps the best candidate
whose existance as a ``missing" resonance can be 
established through $\omega$ photoproduction.

\section*{\bf 5. Conclusion}

In conclusion, we have presented 
a study for $\omega$ photoproduction on the nucleon at low and
 intermediate energies
Our results indicate that s- and u-channel resonances alone are insufficient
at small $t$ and that additional t-channel contributions 
are necessary to describe the large diffractive behavior.
We find that the s- and u-channel resonance contributions are  
important in the large scattering angle region.
The differential cross section alone is 
insufficient to determine s-channel resonance properties in $\omega$
photoproduction, however, polarization observables 
are demonstrated to be very sensitive 
to the presence of individual s-channel resonances. 
It is therefore imperative to perform polarization measurements
in future programs on vector meson 
photoproduction. Our numerical results suggest that properties of the resonance 
$F_{15}(2000)$ could be well determined with precise
$\omega$ photoproduction data.
Given the few free parameters in this 
approach, the agreement with the data, especially 
in the large scattering angle region, is satisfactory. 
Clearly, an improved determination of s-channel resonance properties requires a 
systematic analysis of all photoprduction data.
Due to the small number of free parameters
the quark model approach is an appropriate
tool for this important task.

\section*{Acknowledgments}
Zhenping Li acknowledges the hospitality of the Saclay 
Nuclear Physics Lab, and discussions with B. Saghai. 
We are grateful to F.J. Klein for helpful discussions regarding the data.
This work is supported by the grant 
from Chinese Education Commission and US-DOE grant DE-FG02-95-ER40907.

\newpage
\begin{table}
\caption{The operators appearing in the longitudinal $\pi^0$ exchange expressed 
in terms of helicity amplitudes. ${\hat{\bf k}}$ and ${\hat{\bf q}}$ 
are the unit vectors of ${\bf k}$ and  ${\bf q}$, respectively.
The $d$ functions depend on
the rotation angle $\theta$ between {\bf k} and {\bf q}. 
All other components of $H_{a\lambda_V}$ are zero. $\lambda_f =\pm\frac 12$ 
denotes the helicity of the final state nucleon..}
\protect\label{tab:(1)}
\begin{center}
\begin{tabular}{lcll}
\hline\hline    
Operators&
&$H_{10}(\lambda_f=\frac 12)$&$H_{20}(\lambda_f=\frac 12)$\\
&&$H_{30}(\lambda_f=-\frac 12)$&$H_{40}(\lambda_f=-\frac 12)$\\[1ex]\hline
$(\veps\times{\hat{\bf k}})\cdot{\hat{\bf q}}\vsig\cdot{\hat{\bf q}}$&
&$i\sqrt{2}d^1_{10}  d^1_{10} d^{\frac 12}_{\frac 12\lambda_f} $
&$ i\sqrt{2}d^1_{10}  d^1_{-10} d^{\frac 12}_{-\frac 12\lambda_f } $\\[1ex]
&&$ -id^1_{10}  d^1_{00} d^{\frac 12}_{-\frac 12\lambda_f } $
&$ +id^1_{10}  d^1_{00} d^{\frac 12}_{\frac 12\lambda_f } $\\[1ex]\hline
$(\veps\times{\hat{\bf k}})\cdot{\hat{\bf q}}\vsig\cdot{\hat{\bf k}}$&
&$-id^1_{10}d^{\frac 12}_{-\frac 12 \lambda_f }$
&$id^1_{10}d^{\frac 12}_{\frac 12 \lambda_f }$\\[1ex]\hline
\end{tabular}    
\end{center}
\end{table}

\begin{table}
\caption{ The operators appearing in the transverse $\pi^0$ exchange expressed 
in terms of helicity amplitudes. ${\hat{\bf k}}$ and ${\hat{\bf q}}$
 are the unit vectors of ${\bf k}$ and ${\bf 
q}$, respectively. The $d$ functions depend on the rotation angle 
$\theta$ between {\bf k} and {\bf q}.
$\lambda_V=\pm 1$ denotes the helicity of $\omega$ meson, 
and $\lambda_f=\pm\frac 12$ denotes the helicity of the final nucleon. }
\protect\label{tab:(2)}
\begin{center}
\begin{tabular}{lcll}    
\hline\hline    
Operators&
&$H_{1\lambda_V}(\lambda_f=\frac 12) $
&$H_{2\lambda_V}(\lambda_f=\frac 12) $\\
&&$H_{3\lambda_V}(\lambda_f=-\frac 12) $
&$H_{4\lambda_V} (\lambda_f=-\frac 12)$\\[1ex] \hline

$ (\veps \times\veps_v) \cdot {\hat{\bf q}}\vsig\cdot{\hat{\bf q}}$&
&$i\sqrt{2}\lambda_V d^1_{1\lambda_V}d^1_{10}d^{\frac 12}_{\frac 12\lambda_f}$
&$i\sqrt{2}\lambda_V d^1_{1\lambda_V}d^1_{-10}d^{\frac 12}_{-\frac 12\lambda_f }$\\[1ex]
&&$-i\lambda_V d^1_{1\lambda_V}d^1_{00}d^{\frac 12}_{-\frac 12\lambda_f }$
&$+i\lambda_V d^1_{1\lambda_V}d^1_{00}d^{\frac 12}_{\frac 12\lambda_f }$\\[1ex]\hline
$ (\veps \times \veps_v) \cdot {\hat{\bf q}}\vsig\cdot{\hat{\bf k}}$&
&$-i\lambda_V d^1_{1\lambda_V}d^{\frac 12}_{-\frac 12\lambda_f }$
&$i\lambda_V d^1_{1\lambda_V}d^{\frac 12}_{\frac 12\lambda_f }$\\[1ex]\hline

$ (\veps\times\veps_v) \cdot {\hat{\bf k}}\vsig\cdot{\hat{\bf q}}$&
&$i\sqrt{2} d^1_{1\lambda_V}d^1_{10}d^{\frac 12}_{\frac 12\lambda_f }$
&$i\sqrt{2} d^1_{1\lambda_V}d^1_{-10}d^{\frac 12}_{-\frac 12\lambda_f }$\\[1ex]
&&$-i  d^1_{1\lambda_V}d^1_{00}d^{\frac 12}_{-\frac 12\lambda_f }$
&$+i  d^1_{1\lambda_V}d^1_{00}d^{\frac 12}_{\frac 12\lambda_f }$\\[1ex]\hline
$ (\veps\times\veps_v)\cdot {\hat{\bf k}}\vsig\cdot{\hat{\bf k}}$&
&$-i d^1_{1\lambda_V}d^{\frac 12}_{-\frac 12\lambda_f }$
&$i d^1_{1\lambda_V}d^{\frac 12}_{\frac 12\lambda_f }$\\[1ex]\hline\hline    
\end{tabular}   
\end{center}
\end{table}

\vfill
\newpage
\begin{table}
\caption{ The resonance states 
contributing in $\gamma p\to \omega p$. The star ``*" 
denotes the current determination status of the resonances
as defined in PDG(1996). 
The partial width of resonances decaying 
into $\omega N $
and the photon decay helicity amplitudes $A^\gamma_{\frac 12}$ and 
$A^\gamma_{\frac 32}$ are given by our model. ``-" denotes that 
those states are below the threshold of $\omega$ production 
or the amplitudes decouple for the states.} 
\protect\label{tab:(3)}
\begin{center}
\begin{tabular}{llccccc}    
\hline\hline    
Resonance &$SU(6)$ State&Width&$\sqrt{\Gamma_{\omega}}$
&$\sqrt{\Gamma_{\gamma}}$&$A^\gamma_{\frac 12}$&$A^\gamma_{\frac 32}$\\
&&(MeV)&(MeV$^{-\frac 12}$)&(MeV$^{-\frac 12}$)&&\\[1ex] \hline     
$S_{11}(1535)$****&$N(^2P_M)_{\frac 12^-}$&150&-&2.67&+172.4&-\\[1ex] 
$D_{13}(1520)$****&$ N(^2P_M)_{\frac 32^-}$&120&-&0.31&-51.8&+122.5\\[1ex]    
$P_{13}(1720)$****&$N(^2D_S)_{\frac 32^+}$&150&-&0.61&-113.6&+38.0\\[1ex]    
$F_{15}(1680)$****&$N(^2D_S)_{\frac 52^+}$&130&-&0.10&0&+78.5\\[1ex]    
$P_{11}(1440)$****&$N(^2S^\prime_S)_{\frac 12^+}$&350&-&0.03&-37.4&-\\[1ex]    
$P_{11}(1710)$***&$N(^2S_M)_{\frac 12^+}$&100&-&0.12&-37.7&-\\[1ex]    
$P_{13}(1900)$**&$N(^2D_M)_{\frac 32^+}$&400&1.40&0.48&+90.4&-25.8\\[1ex]    
$F_{15}(2000)$**&$N(^2D_M)_{\frac 52^+}$&200&5.12&0.09&+28.6&-49.9\\[1ex]
\hline    
\end{tabular}    
\end{center}
\end{table}

\newpage
\section*{Figure Captions:}
\begin{enumerate}
\item The differential cross sections (solid curves) for $\gamma p\to\omega p$
at $E_\gamma=$1.225, 1.45, 1.675 
and 1.915 GeV. The data come from Ref.~\cite{elsa}. 
The $\pi^0$ exchanges are shown by the dashed curves 
and the contributions from s- and u-channel 
exclusively are shown by the dotted curves.
In (c), the dot-dashed curve represents the 
differential cross section without 
contributions from the resonance $F_{15}(2000)$.

\item The four single-spin polarization observables in $\gamma p\to\omega p$
are given by the solid curves at $E_\gamma=1.675$GeV. The dotted curves 
correspond to the asymmetries without the resonance $F_{15}(2000)$, 
while the dashed curves to those without the resonance $P_{13}(1900)$.

\item The total cross section of $\gamma p\to\omega p$ are fitted by the
solid curve with $\pi^0$ exchange taken into account. The dotted curve
describes the pure contributions from s- and u-channel. The data are
from~\cite{elsa}(triangle), \cite{ABHMC} and other 
experiments(square).
\end{enumerate}


\begin{thebibliography}{99}
\bibitem{NRQM} N. Isgur and G. Karl, Phys. Letts. {\bf 72B}, 109(1977); Phys. Rev. {\bf D23}, 
817(1981); R. Koniuk and N. Isgur, Phys. Rev. {\bf D21}, 1868(1980).
\bibitem{capstick} S. Capstick, and W. Roberts, Phys. Rev. {\bf D49}, 4570(1994).
\bibitem{elsa} F. Klein et al., to appear in the Proceedings of the GW/TJNAF Workshop 
on $N^*$  Physics, (Washington, D.C.) 1997, F.J. Klein, Ph.D. thesis, University of Bonn (1996).
\bibitem{cebaf} TJNAF experimental proposals
E94-109 (P. Cole, R. Whitney, Co-spokesmen);
E91-024 (V. Burkert, H. Funsten. M. Manley, B. Mecking, Co-spokesmen).
\bibitem{vector} Q. Zhao, Z.P.Li and C. Bennhold, nucl-th/9711061.
\bibitem{duality} R. Dolen, D. Horn, and C. Schmid, Phys. Rev. {\bf 166}, 1768(1966).
\bibitem{david}  J. C. David, {\it et al.} Phys. Rev. {\bf C53}, 2613(1996).
\bibitem{FrimanSoyeur} B. Friman and M. Soyeur, Nucl. Phys. {\bf A100}, 477(1996).
\bibitem{tabakin} M. Pichowsky, \c{C}. \c{S}avkl\i, F. Tabakin, Phys. Rev. 
{\bf C53}, 593 (1996).
\bibitem{eta} Zhenping Li, Phys. Rev. {\bf D52}, 4961 (1995).
\bibitem{pdg} E.J.Weinberg, {\it et al}, Phys. Rev. {\bf D54}, 1(1996).
\bibitem{moor} R.G.Moorhouse,Phys. Rev. Lett. {\bf 16}, 772(1966).
\bibitem{soding}  P. S\"oding, Phys. Lett. {\bf 19}, 702(1965).
\bibitem{pomeron} M. Pichowsky, ``Nonperturbative Quark Dynamics in 
diffractive processes", Ph.D. Dissertation, The University Of Pittsburgh,
(1996); and references therein.  
\bibitem{pseudoscalar} Zhenping Li, Ye Hongxing and Lu Minghui, Phys. Rev. {\bf C56}, 1099 (1997).
\bibitem{Compton} Zhenping Li, Phys. Rev. {\bf D48}, 3070(1993).
\bibitem{wolf} G. Wolf, Proc. 1971 Int. Symp. on Electron and Photon Interactions at High 
Energies, Cornell University, Ithaca(USA), August 23-27,1971,p.190.
\bibitem{capstickcompton} S. Capstick and B. D. Keister, Phys. Rev. {\bf D46},
84 (1992).
\bibitem{ABHMC} Aachen-Berlin-Bonn-Hamburg-Heidelberg-M\"unchen Collabration, 
Phys. Rev. {\bf 175}, 1669(1968).
\bibitem{data} Crouch {\it et al}, Phys. Rev. {\bf 155}, 1468(1967);
Y. Eisenberg {\it et al}, Phys. Rev. {\bf D5}, 15(1972);
Y. Eisenberg {\it et al}, Phys. Rev. Lett. {\bf 22}, 669(1969);
W. C. Barber {\it et al}, Z. Phys. {\bf C26}, 343(1984);
J. Ballam {\it et al}, Phys. Rev. {\bf D7}, 3150(1973);
W.Struczinski {\it et al}, Nucl. Phys. {\bf B108}, 45(1976).
\bibitem{thesis} Zhenping Li, ``Photo- and Electroproduction of 
Baryon Resonances in A Potential Quark Mode", Ph.D. Dissertation, 
The University of Tennessee, (1991).
\bibitem{program} V. D. Burkert, The N* Program at CEBAF, CEBAF-PR-92-021.
\bibitem{bijan} B. Saghai and F. Tabakin, Phys. Rev. {\bf C55}, 917(1995).
\end{thebibliography}
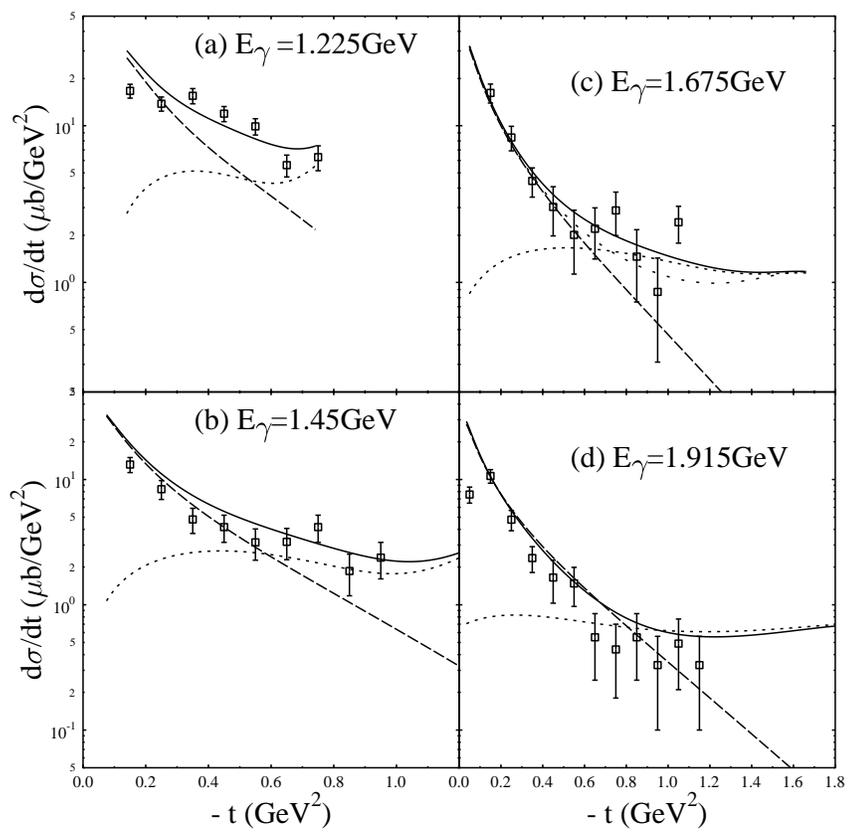
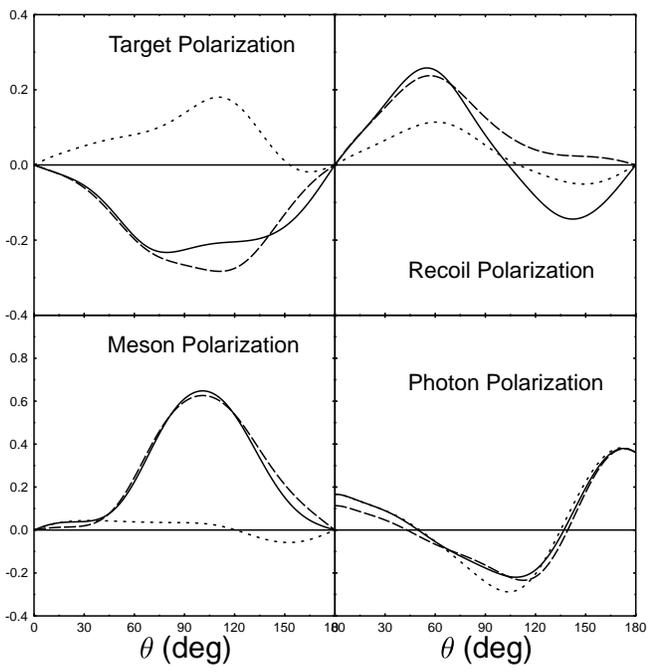
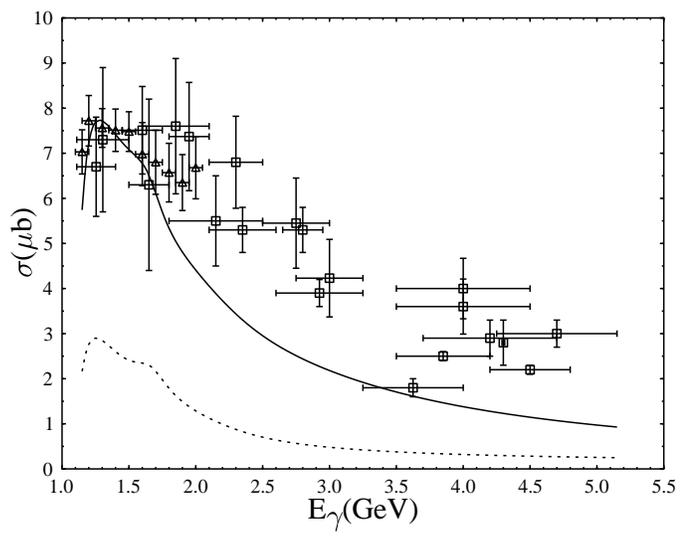
\end{document}